\documentclass[twocolumn,noshowpacs,preprintnumbers,amsmath,amssymb]{revtex4}

\usepackage{graphicx}
\usepackage{dcolumn}
\usepackage{bm}
\usepackage{color}
\usepackage[draft,english]{fixme}

\begin{document}

\title{Neutrino Absorption in the Earth, Neutrino Cross-Sections, and New Physics}

\author{Spencer R. Klein} \affiliation{Lawrence
Berkeley National Laboratory, Berkeley CA 94720 USA,  and the University of California, Berkeley, CA 94720 USA}
\author{Amy Connolly} \affiliation{The Ohio State University, Columbus, OH 43210 USA, and the Center for Cosmology and Astroparticle Physics (CCAPP), Columbus, OH 43210 USA.}
 

\begin{abstract}

\end{abstract}

\maketitle

\begin{figure}[htbp]
\begin{center}
\includegraphics[width=0.5\textwidth]{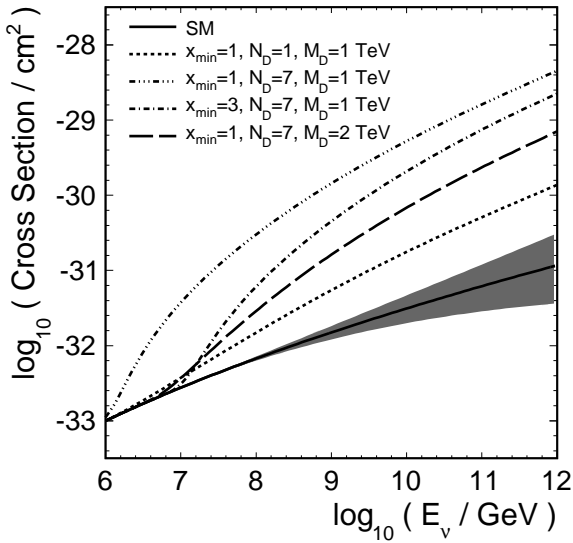}
\caption{The neutrino interaction cross-section for the standard model, plus several models that incorporate additional dimensions.  From Ref. \cite{Connolly:2011vc}.
\label{fig:sigma}
}
\end{center}
\end{figure}

Large neutrino telescopes can measure the neutrino-nucleon cross-section by studying neutrino absorption in the Earth \cite{Hooper:2002yq,Borriello:2007cs}.   At high energies, this cross-section is sensitive to new physics.  In particular, if there are additional rolled-up dimensions, then the cross-section will increase sharply at an energy corresponding to the inverse size of the extra dimension(s).    Figure \ref{fig:sigma} shows the neutrino-nucleon cross-sections calculated for the Standard Model, plus several models with additional dimensions \cite{Connolly:2011vc}.  Other types of new physics can produce similar effects.  
For example, the presence of leptoquarks could produce a similar increase in the cross-section \cite{Romero:2009vu}.  

\begin{figure}[tb]
\begin{center}
\includegraphics[width=0.5\textwidth]{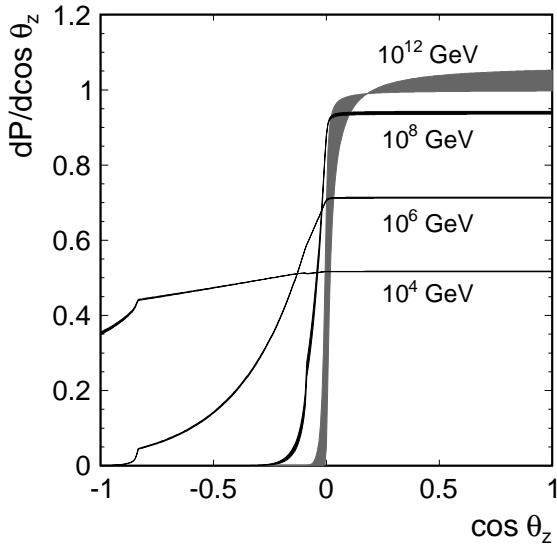}
\caption{The zenith angle distribution for high-energy neutrinos of different energies, for Standard Model cross sections.  At energies above $10^{17}$ eV, absorption in the Earth limits the flux of upward-going neutrinos, except near the horizon.  From Ref. \cite{Connolly:2011vc}.
\label{fig:angles}
}
\end{center}
\end{figure}

Neutrino absorption becomes an effective technique for measuring the cross-section at neutrino energies above about 50 TeV, the energy at which absorption (assuming the Standard Model cross-sections)  begins to reduce the flux of vertically upward-going neutrinos, altering the zenith angle distribution.  Figure \ref{fig:angles} shows the expected zenith angle distribution for neutrinos with energies between $10^{13}$ and $10^{21}$ eV.   Lower energy neutrinos can be used as a normalization, to check the angular acceptance of the detector, and to calibrate for the small zenith angle dependence in the atmospheric neutrino flux. 

At neutrino energies much above $10^{17}$ eV, the cross-sections depend significantly on parton distributions at Bjorken$-x$ and $Q^2$ values beyond the reach of HERA data, so extrapolations are required to predict the cross-sections.  LHC data can be used to contrain the parton distributions, but, even at current experiments like IceCube, the neutrino energies are 100 times higher than are accessible at accelerators.  So, suprises are certainly possible, especially for neutrino-philic new physics. 

Two classes of analyses are foreseen.  In the short term, IceCube \cite{IceCube} and proposed Km3Net \cite{Sapienza:2011zzb} optical Cherenkov detectors can study the absorption of atmospheric neutrinos.  These are a well understood source; the difficulty is that the flux drops rapidly with energy, and so adequate statistics are only available in a fairly narrow energy range (perhaps up to 100 TeV).  If the flux of prompt neutrinos is high enough, they may enable the range to be extended slightly.  These analyses are also somewhat sensitive to the composition of the Earth's core \cite{GonzalezGarcia:2007gg}.   However, the core composition is already fairly well known, and a very large data set is needed to compete with the geophysical measurements.  The current core uncertainties are less important to this analysis as the neutrino cross-section uncertainties.  

In the longer term, the proposed ARA \cite{Allison:2011wk} and ARIANNA \cite{Klein:2012bu} radio Cherenkov detectors may measure the neutrino flux at much higher energies, above $10^{17}$  eV.  These detectors observe the radio pulse from the electromagnetic and hadronic showers produced by neutrino interactions in Antarctic ice.  They rely on astrophysical neutrinos for the measurement.  Rate calculations are based on Greissen-Zatsepin-Kumino (GZK) neutrinos,  produced when cosmic microwave background radiation photons excite cosmic-ray protons with energies above $4\times 10^{19}$ eV to a $\Delta^+$ resonance.  The flux should peak around $10^{19}$ eV \cite{Engel:2001hd}.   This flux is well calculated and should be present as long as these high energy cosmic-rays are protons.  For heavier nuclei, the energy per nucleon is lower, and the neutrino production is much lower.

At these energies, for even the Standard Model cross-sections, the Earth is almost completely opaque to neutrinos, so the cross-section measurement relies on the observation of neutrinos near the horizon.    Figure \ref{fig:angles} shows the angular distribution for neutrinos of different energies.  At energies above $10^{17}$ eV, the flux extends only a few degrees below the horizon, so this measurement requires good angular resolution.  ARA simulations predict a mean resolution of about 6$^\circ$ \cite{Allison:2011wk}, while ARIANNA simulations give a mean zenith angle resolution of 2.9$^\circ$, with a similar azimuthal resolution.    Both experiments expect to observe of order 10 events per year with a full detector (37 ARA stations, 900 ARIANNA stations).   The GZK neutrino energy spectrum extends up to about $10^{19}$ eV, corresponding to a $\nu$-nucleon center-of-mass energy of 140 TeV.   

Because these analyses measure neutrino disappearance, they cannot easily differentiate between the charged and neutral current cross-sections; in the latter case, the neutrino survives, but at a lower energy.  So, the analyses must make some assumptions about the relative cross-sections; one can assume that the two cross-sections scale linearly from the standard-model assumption.  

CERN's Large Hadron Collider (LHC) can make similar measurements, at, by 2015, proton-proton center of mass energies up to 14 TeV, about 1/10 the $\nu$-nucleon energy accessible with GZK neutrinos.  The LHC will have a much higher luminosity, so, for slowly rising cross-sections, will have higher sensitivity.     However, for scenarios where the cross-section rises rapidly, or for neutrino-philic new physics, UHE neutrinos may be more sensitive.

We thank the organizers of the "Snowmass" process for the opportunity to present our thoughts.  Connolly would like to acknowledge support from the Ohio State University and CCAPP, and the U.S. National Science Foundation under grant  PHY-1255557.  Kleins acknowledges support from the U.S. National Science Foundation under grant 0653266 and the U.S. Department of Energy under contract number DE-AC-76SF00098.


\begin{thebibliography}{99}

\def\etal{{\it et al.}}
\bibitem{Hooper:2002yq} 
  D.~Hooper,
  Phys.\ Rev.\ D {\bf 65}, 097303 (2002)
  [hep-ph/0203239].

\bibitem{Borriello:2007cs}
  E.~Borriello, A.~Cuoco, G.~Mangano, G.~Miele, S.~Pastor, O.~Pisanti and P.~D.~Serpico,
  Phys.\ Rev.\ D {\bf 77} (2008) 045019
  [arXiv:0711.0152 [astro-ph]].

\bibitem{Connolly:2011vc} 
  A.~Connolly, R.~S.~Thorne and D.~Waters,
  Phys.\ Rev.\ D {\bf 83}, 113009 (2011).

\bibitem{Romero:2009vu} 
  I.~Romero and O.~A.~Sampayo,
  JHEP {\bf 0905}, 111 (2009)
  [arXiv:0906.5245 [hep-ph]].

\bibitem{GonzalezGarcia:2007gg} 
  M.~C.~Gonzalez-Garcia, F.~Halzen, M.~Maltoni and H.~K.~M.~Tanaka,
  Phys.\ Rev.\ Lett.\  {\bf 100}, 061802 (2008).

\bibitem{IceCube} S. Miarecki, presented at the Intl. Neutrino School, July 18-30, 2011, Cartigny, Switzerland.  Available at
http://indico.cern.ch/getFile.py/access?contribId=55\ \&sessionId=0\&resId=1\&materialId=slides\&confId=147174;
F.~Halzen and S.~R.~Klein,  Rev.\ Sci.\ Instrum.\  {\bf 81}, 081101 (2010).

\bibitem{Sapienza:2011zzb} 
  P.~Sapienza [KM3NeT Collaboration],
  Nucl.\ Phys.\ Proc.\ Suppl.\  {\bf 212-213}, 134 (2011).

\bibitem{Allison:2011wk} 
  P.~Allison {\it et al.},
  Astropart.\ Phys.\  {\bf 35}, 457 (2012).

\bibitem{Klein:2012bu} 
  S.~R.~Klein [ARIANNA Collaboration],
  arXiv:1207.3846;
 L.~Gerhardt {\it et al.},
  Nucl.\ Instrum.\ Meth.\ A {\bf 624}, 85 (2010).

\bibitem{Engel:2001hd} 
  R.~Engel, D.~Seckel, T.~Stanev and ,
  Phys.\ Rev.\ D {\bf 64}, 093010 (2001)
  [astro-ph/0101216].

\end{thebibliography}
\end{document}